\documentclass[epj,final]{svjour}

\usepackage{latexsym}
\usepackage{url}
\usepackage{amsfonts}
\usepackage{amssymb}

\RequirePackage{graphicx}

\begin{document}
\title{The cosmological constant derived via galaxy groups and clusters}
\author{V.G. Gurzadyan\inst{1,2}, 
A. Stepanian\inst{1}
}                     
%
%
\institute{Center for Cosmology and Astrophysics, Alikhanian National Laboratory and Yerevan State University, Yerevan, Armenia \and
SIA, Sapienza Universita di Roma, Rome, Italy}
\date{Received: date / Revised version: date}
%

\abstract{
The common nature of dark matter and dark energy is argued in \cite{G} based on the approach that the cosmological constant 
$\Lambda$ enters the weak-field General Relativity following from Newton theorem on the "sphere-point mass" equivalency \cite{GS1}. Here we probe the $\Lambda$-gravity description of dark matter in galaxy systems, from pairs up to galaxy clusters using the data of various sources, i.e. of Local Supercluster galaxy surveys, gravity lensing and Planck satellite. The prediction that the cosmological constant has to be the lower limit for the weak-field $\Lambda$ obtained from galaxy systems of various degree of virialization is shown to be supported by those observations. The results therefore support the $\Lambda$-gravity nature of dark matter in the studied systems, implying that the positivity of the cosmological constant might be deduced decades ago from the dynamics of galaxies and galaxy clusters far before the cosmological SN surveys.        
} 
%
\maketitle
\section{Introduction}

A number of approaches are considered to reveal the nature of dark matter, including prediction of exotic particles and modified gravity models; for review see \cite{Bert}.  One of the recent approaches \cite{G,GS1} is based on the General Relativity (GR) with a modified weak field limit following from the Newton theorem on equivalency of gravity of sphere and of point mass. That approach enables the common description of dark matter and dark energy, where $\Lambda$ acts as a universal constant defining the GR and its weak field limit, along with the gravitational constant $G$  \cite{GS1}. An observational test for that approach is suggested involving the effect of gravity lensing \cite{GS2}. 

The notable point following from the Newton theorem is that, on the one hand, $\Lambda$ acts as a cosmological constant describing the expansion of the Universe, on the other hand, the same it defines the weak field gravity proper for the dynamics of galaxies, galaxy groups and clusters, i.e. on the distance scales where the Hubble flow cannot be valid. As shown in \cite{G} the value of the cosmological constant indeed is supported by the parameters of galactic halos and of a sample of galaxy groups.

Here we continue to address the principal issue on the $\Lambda$-nature of dark matter considering the data on galaxy systems, from galaxy pairs to galaxy clusters. Among the used samples are data on galaxy groups of Local Supercluster, of galaxy clusters obtained both with gravity lensing and by Planck satellite. Assuming that the dynamical structure of the considered galaxy systems is governed by the modified Newton law \cite{G}, we obtain $\Lambda$ which appear limited from below by the value of cosmological constant. In other words, the value of the cosmological constant is extracted from galaxy systems and not from cosmology.

\section{The $\Lambda$ constant and galaxy systems}

The Newton theorem on ``sphere-point" equivalency enables one to arrive at the GR metric \cite{G,GS1}
\begin {equation} 
g_{00} = 1 - \frac{2 G m}{r c^2} - \frac{\Lambda r^2}{3};\,\,\, g_{rr} = 1 + \frac{2 G m}{r c^2} + \frac {\Lambda r^2}{3}.
\end {equation} 
This implies that the weak field limit for GR as modified Newton law involves two constants (for the potential)
\begin{equation}\label{form}
\phi(r)= C_1 r^{-1} + C_2 r^2,
\end{equation}   
where $C_1$ is assigned to $G$ and $C_2$ to $\Lambda$ (within numerical factors).

Eq.(1) was known before as Schwarzschild -- de Sitter metric \cite{R}, where the constant $\Lambda$ was introduced by Einstein to describe the static cosmological model. 

Within our approach $\Lambda$ is emerging from the general function satisfying the Newton's theorem and hence naturally emerges in weak field GR, and that correspondence can be represented via isometry groups. Namely, the Lorentz group O(1,3) acts as stabilizer subgroup of isometry group of 4D maximally symmetric Lorentzian geometries \cite{GS1}. The results are shown in Table \ref{group} and for all these geometries the Lorentzian spheres can be considered as points. Consequently, depending on the sign of $\Lambda$, we will have the following non-relativistic limits  
\begin{equation}\label{nrel}
O(1,4) \to (O(3) \times O(1,1)) \ltimes R^6,\quad O(2,3) \to (O(3) \times O(2)) \ltimes R^6, \quad IO(1,3) \to (O(3) \times R) \ltimes R^6,
\end{equation}
for all three cases, since the spatial algebra is Euclidean i.e. 
\begin{equation}
E(3)=R{}^{3} \rtimes O(3).
\end{equation}
Hence, for all three cases introduced above the O(3) is the stabilizer group for spatial geometry. This conclusion, in its turn, can be considered as the Newton's theorem in the language of group theory.

\begin{table}
\caption{Background geometries for vacuum solutions}\label{group}
\centering
\begin{tabular}{ |p{2cm}||p{2.4cm}|p{2.4cm}|p{1.5cm}| }
\hline

& Spacetime&Isometry group\\
\hline
$\Lambda > 0$ &de Sitter  &O(1,4)\\
$\Lambda = 0$ & Minkowski  & IO(1,3)\\
$\Lambda <0 $ &Anti de Sitter  &O(2,3)\\
\hline
\end{tabular}
\end{table}

Then, as shown in \cite{G}, the weak field $\Lambda$ extracted from galactic halo and galaxy group data quantitatively coincides with the $\Lambda$ obtained from cosmological data. 

The value of $\Lambda$ as of the cosmological constant is currently obtained by several observational methods, the one obtained by Planck data yields \cite{P} 
\begin{equation}
\Lambda_{PL}=1.11\,\,  10^{-52} m^{-2}. 
\end{equation}
For $\Lambda$-modified Newtonian gravity Eq.(2) the following relation for virialized systems was derived \cite{G}
\begin{equation}\label{Es}
\Lambda= \frac{3 \sigma^{2}}{2 c^2 R^2},
\end{equation}
where $R$ is system's radius and $\sigma$ is the velocity dispersion.

To test the conclusion on $\Lambda$-nature of dark matter in galaxy systems, below we use Eq.(6) for the analysis of data samples on galaxy pairs, galaxy groups and galaxy clusters.

\subsection{Galaxy pairs}

To probe Eqs.(1),(2) for galaxy systems we start with galaxy pairs. Here, using the weak lensing data by Gonzalez et al \cite{GB}, we estimate the value of $\Lambda$ from Eq.(6) as shown in Tables 2 and 3 for given classes of subsamples. Estimations are done both for Singular Isothermal sphere (SIS) and Navarro-Frenk-White (NFW) models as described in details in \cite{GB}, $M_{200}$  corresponds the mass (for the given model) within radius $R_{200}$, when the mean density equals 200 critical density values of the universe. 

\begin{table}[h]
\caption{SIS profile}\label{tab1}
\centering
\begin{tabular}{ |p{3.6cm}|p{1.5cm}|p{2.4 cm}||p{1.5cm}| }
\hline

Subsamples & $\sigma$ (Km/s) & M${}_{200} ($ M${}_{\odot}$)&$\Lambda$ ($m^{-2}$)  \\ \hline
\hline
Total Sample& 223 $\pm {24}$  &  (6.9 $\pm {2.2}$)E12 &5.64E-51\\ \hline
Non Interacting& 200 $\pm {38}$ & (5.0 $\pm {3.0}$)E12 &5.62E-51\\ \hline
Interacting Pairs & 237 $\pm {29}$  &  (8.3 $\pm {3.0}$)E12 &5.63E-51\\ \hline
Red Pairs&264 $\pm {28}$ &  (11.4 $\pm {3.7}$)E12 &5.65E-51\\ \hline
Blue Pairs& 167 $\pm {45}$  &  (2.9 $\pm {2.4}$)E12 &5.64E-51\\ \hline
Higher Luminosity Pairs& 278 $\pm {27}$ & (13.2 $\pm {3.8}$)E12 &5.69E-51\\ \hline
Lower Luminosity Pairs& 149 $\pm {50}$ &  (2.0 $\pm{2.0}$)E12 &5.75E-51\\ \hline
\hline

Average  &      &     &  5.66E-51 \\ \hline
    St. deviation	   &      &     &  4.21E-53 \\ \hline
		\hline

\end{tabular}
\end{table}

\begin{table}[h]
\caption{NFW profile}\label{tab2}
\centering
\begin{tabular}{ |p{3.6cm}|p{1.5cm}|p{2.4 cm}||p{1.5cm}| }
\hline

Subsamples & R${}_{200}$ (Mpc) & M${}_{200} ($ M${}_{\odot}$)&$\Lambda$ ($m^{-2}$)  \\ \hline
		\hline
Total Sample& 0.30 $\pm{0.03}$  &  (7.1 $\pm {2.1}$)E12 &2.00E-50\\ \hline
Non Interacting& 0.27 $\pm{0.05}$ & (5.2 $\pm {2.8}$)E12 &2.01E-50\\ \hline
Interacting Pairs & 0.32 $\pm{0.03}$ & (8.0 $\pm {2.9}$)E12 &1.86E-50\\ \hline
Red Pairs&0.36 $\pm{0.03}$  & (12.1 $\pm {3.6}$)E12 &1.97E-50\\ \hline
Blue Pairs& 0.22 $\pm{0.06}$  &   (2.7 $\pm {2.1}$)E12 &1.93E-50\\ \hline
Higher Luminosity Pairs& 0.36 $\pm{0.03}$ & (12.7 $\pm {3.8}$)E12 &2.07E-50\\ \hline
Lower Luminosity Pairs& 0.19  $\pm{0.07}$ & (1.7 $\pm{1.8}$)E12 &1.89E-50\\ \hline
\hline

Average  &      &     &  1.96E-50 \\ \hline
    St. deviation	   &      &     &  6.77E-52 \\ \hline
		\hline

\end{tabular}
\end{table}

\subsection{Galaxy groups}

Data on galaxy groups as of systems containing 3 and more galaxies situated within the Local Supercluster, namely, of the Leo/Cancer region and Bootes strip are obtained by Karachentsev et al \cite{LC,BS}; the analysis of their data on groups of the Hercules-Bootes region \cite{HB} is given in \cite{G}. Tables 4 and 5 contain the observational data i.e. the velocity dispersion of galaxies $\sigma$ and the harmonic average radius $R_h$ of the groups denoted by their brightest galaxy listed in the first column. The last column contains $\Lambda$ estimated by Eq.(6).

\begin{table}[!h]
\caption{Galaxy groups of Leo/Cancer region}\label{tab4}
\centering
\begin{tabular}{ |p{1.8cm}|p{1.8cm}|p{1.2 cm}||p{1.5cm}| }
\hline

Group    &  $\sigma_V(km/s^{-1})$ & $R_h(kpc)$ & $\Lambda (m^{-2}$) \\ \hline		
		\hline
NGC2648 &  55 & 128  & 3.24E-51  \\ \hline
NGC2775 &  89 & 296  & 1.59E-51  \\ \hline
NGC2894 &  50 & 458  & 2.09E-52  \\ \hline
NGC2962 &  53 & 161  & 1.90E-51  \\ \hline
NGC2967 &  62 & 507  & 2.63E-52  \\ \hline
UGC5228 &  40 & 188  & 7.95E-52  \\ \hline 
NGC3023 &  21 &  35  & 6.32E-51  \\ \hline
NGC3020 &  45 &  44  & 1.84E-50  \\ \hline
NGC3049 &  15 & 144  & 1.91E-52  \\ \hline 
UGC5376 &  66 & 253  & 1.20E-51  \\ \hline
NGC3166 &  44 & 126  & 2.14E-51  \\ \hline
NGC3227 &  74 & 128  & 5.87E-51  \\ \hline
NGC3338 &  50 & 112  & 3.50E-51  \\ \hline
NGC3379 &  193& 191  & 1.79E-50  \\ \hline
NGC3423 &  21 & 570  & 2.38E-53  \\ \hline
NGC3521 &  37 & 132  & 1.38E-51  \\ \hline
NGC3596 &  42 &  41  & 1.84E-50  \\ \hline
NGC3607 &  115& 471  & 1.05E-51  \\ \hline
NGC3626 &  86 & 187  & 3.72E-51  \\ \hline
NGC3627 &  136& 201  & 8.04E-51  \\ \hline
NGC3640 &  134& 252  & 4.97E-51  \\ \hline
NGC3686 &  91 & 175  & 4.75E-51  \\ \hline
NGC3810 &  43 & 360  & 2.51E-52  \\ \hline
\hline
    Average  &      &     &  4.62E-51 \\ \hline
    St. deviation	   &      &     &  5.70E-51 \\ \hline
		\hline
\end{tabular}
 \end{table}

\begin{table}[!h]
\centering
\caption{Galaxy groups of Bootes strip region}
\begin{tabular}{|p{1.8cm}|p{1.8cm}|p{1.2 cm}||p{1.5cm}|}
\hline

    Group    &  $\sigma_V(km/s^{-1})$ & $R_h(kpc)$ & $\Lambda (m^{-2}$) \\ \hline		
  	\hline
N4900  & 36 & 116 & 1.69E-51  \\ \hline
N5248  & 38 & 151 & 1.11E-51  \\ \hline
N5363  & 114& 165 & 8.39E-51  \\ \hline
N5506  & 23 &  35 & 7.59E-51  \\ \hline
N5566  & 103& 196 & 4.85E-51  \\ \hline
N5638  & 74 & 203 & 2.33E-51  \\ \hline
P51971 & 10 & 100 & 1.76E-52  \\ \hline
IC1048 & 83 & 150 & 5.38E-51  \\ \hline
N5746  & 107& 296 & 2.30E-51  \\ \hline
N5775  & 87 & 120 & 9.23E-51  \\ \hline
N5792  & 48 & 290 & 4.81E-52  \\ \hline
N5838  & 53 & 210 & 1.12E-51  \\ \hline
N5846  & 228& 415 & 5.30E-51  \\ \hline 
\hline
    Average  &      &     &  3.84E-51 \\ \hline
    St. deviation	   &    &     &  3.01E-51 \\ \hline
		\hline
\end{tabular}
\end{table}

\subsection{Galaxy clusters}

We now turn to galaxy clusters, the higher scale structure in the hierarchy of galaxy systems. We use the data both on cluster gravity lensing and supernova survey with HST (CLASH) \cite{CL}. In the Table 6 we exhibit the data for all 19 clusters of CLASH survey along with the results of estimation of $\Lambda$ using Eq.(6).
\begin{table}[!h]
\caption{CLASH Survey}
\centering
\begin{tabular}{ |p{3.6cm}|p{1.2cm}|p{2.4 cm}||p{1.5cm}| }
\hline

Cluster& R${}_{vir}$ (Mpc) & M${}_{vir}$ ( M${}_{\odot}$) & $\Lambda$ ($m^{-2}$) \\ \hline
		\hline
Abell 383& 1.86  & (1.04 $\pm {0.07}$)E15 &  1.23E-50\\ \hline
Abell 209& 1.95 & (1.17 $\pm {0.07}$)E15 & 1.20E-50\\ \hline
Abell 2261& 2.26  &  (1.76 $\pm {0.18}$)E15 & 1.16E-50\\ \hline
RXJ2129+0005&1.65 & (0.73 $\pm {0.07}$)E15 &  1.24E-50\\ \hline
Abell 611& 1.79 &  (1.03 $\pm {0.07}$)E15 & 1.37E-50\\ \hline
MS2137-2353& 1.89 & (1.26 $\pm {0.06}$)E15 & 1.42E-50\\ \hline
RXCJ2248-4431& 1.92 & (1.40 $\pm{0.12}$)E15 & 1.51E-50\\ \hline
MACSJ1115+0129&1.78 & (1.13 $\pm{0.10}$)E15 &1.52E-50\\ \hline
MACSJ1931-26& 1.61 &  (0.83 $\pm{0.06}$)E15 &1.51E-50\\ \hline
RXJ1532.8+3021& 1.47 & (0.64 $\pm{0.09}$)E15 & 1.53E-50\\ \hline
MACSJ1720+3536& 1.61 & (0.88 $\pm{0.08}$)E15 &1.60E-50\\ \hline
MACSJ0429-02& 1.65 &  (0.96 $\pm{0.14}$)E15 &1.63E-50\\ \hline
MACSJ1206-08& 1.63 &  (1.00 $\pm{0.11}$)E15 &1.76E-50\\ \hline
MACSJ0329-02& 1.54 & (0.86 $\pm{0.11}$)E15 &1.79E-50\\ \hline
RXJ1347-1145& 1.80 & (1.35 $\pm{0.19}$)E15 & 1.76E-50\\ \hline
MACSJ1311-03& 1.28 & (0.53 $\pm{0.04}$)E15 & 1.92E-50\\	\hline
MACSJ1423+24& 1.34 & (0.65 $\pm{0.11}$)E15 &2.06E-50\\	\hline
MACSJ0744+39& 1.33 & (0.79 $\pm{0.04}$)E15 &2.56E-50\\ \hline
CLJ1226+3332& 1.57& (1.72 $\pm{0.11}$)E15 &3.38E-50\\ \hline
\hline

	Average  &      &     &  1.69E-50 \\ \hline
    St. deviation	   &    &     &  5.16E-51 \\ \hline
		\hline

\end{tabular}
\end{table}

Besides the CLASH data we use also data of 3 Planck clusters from \cite{PC} and represent the estimated $\Lambda$ (Table 7).

\begin{table} [!h]
\caption{Planck Clusters}\label{tab6}
\centering
\begin{tabular}{ |p{3.3cm}|p{1.8cm}|p{2.1 cm}||p{1.5cm}| }
\hline

Clusters &  $\sigma$ (Km/s) & M${}_{200} ($ M${}_{\odot}$)&$\Lambda$ ($m^{-2}$)  \\ \hline
\hline
PSZ1 G109.88+27.94 & 1800 $\pm{200}$  &  (44 ${}_{-13}^{+16}$)E14 &4.96E-51\\ \hline
PSZ1 G139.61+24.20 & 800 $\pm{100}$ & (6.3 ${}_{-2.1}^{+2.7}$)E14 &3.58E-51\\	\hline
PSZ1 G186.98+38.66 & 1100 $\pm{200}$ & (14.5 ${}_{-6.5}^{+9.4}$)E14 &3.88E-51\\ \hline
\hline
Average  &      &     &  4.14E-51 \\ \hline
    St. deviation	   &    &     &  5.93E-52 \\ \hline
		\hline

\end{tabular}
\end{table}
Finally, we estimate the $\Lambda$ for a sample of Local Cluster Substructure Survey (LoCuSS) clusters \cite{LoCuSS}, as represented in Table 8.
\begin{table} [h]
\caption{LoCuSS Clusters}\label{tab7}
\centering
\begin{tabular}{ |p{2.8cm}|p{1.5cm}|p{2.4cm}||p{1.5cm}| }
\hline

Cluster& R${}_{200}$ ($h^{-1}_{100}Mpc)$ & M${}_{200}$ ( M${}_{\odot}$) & $\Lambda$ ($m^{-2}$) \\ \hline
\hline
Abell 586& 1.2 $\pm{0.2}$  & (5.1 $\pm {2.1}$)E14 &  2.25E-50\\ \hline
Abell 611& 1.1 $\pm{0.1}$ & (4.0 ${}^{+0.7}_{-0.8}$)E14& 2.29E-50\\ \hline
Abell 621& 1.2 ${}^{+0.2}_{-0.1}$  &  (4.8 ${}^{+1.7}_{-1.8}$)E14 & 2.11E-50\\ \hline
Abell 773&1.1 $\pm{0.1}$ & (3.6 $\pm {1.2}$)E14 &  2.06E-50\\ \hline
Abell 781& 1.1 $\pm{0.1}$ &  (4.1 $\pm {0.8}$)E14  & 2.34E-50\\ \hline
Abell 990& 0.9 $\pm{0.1}$ & (2.0 ${}^{+0.4}_{-0.1}$)E14 & 2.09E-50\\ \hline
Abell 1413& 1.1 $\pm{0.1}$ & (4.0 $\pm{1.0}$)E14 & 2.29E-50\\ \hline
Abell 1423&0.9 $\pm{0.1}$ & (2.2 $\pm{0.8}$)E14 &2.30E-50\\ \hline
Abell 1758a& 1.1 $\pm{0.1}$ &  (4.1 ${}^{+0.7}_{-08}$)E14 &2.34E-50\\ \hline
Abell 1758b& 1.1 $\pm{0.2}$ & (4.4 $\pm{1.9}$)E14 & 2.52E-50\\ \hline
Abell 2009& 1.2 $\pm{0.1}$ & (4.6 $\pm{1.5}$)E14 &2.03E-50\\ \hline
Abell 2111& 1.1 $\pm{0.1}$ &  (4.2 $\pm{0.9}$)E14 &2.40E-50\\ \hline
Abell 2146& 1.2$\pm{0.1}$ &  (5.0 $\pm{0.7}$)E14 &2.20E-50\\ \hline
Abell 2218& 1.3$\pm{0.1}$ & (6.1 $\pm{0.9}$)E14 &2.11E-50\\ \hline
RXJ0142+2131& 1.0$\pm{0.1}$ & (3.7 ${}^{+1.1}_{-1.2}$)E14 & 2.82E-50\\ \hline
RXJ1720+2638& 0.9$\pm{0.1}$ & (2.0 $\pm{0.4}$) E14 & 2.09E-50\\ \hline
\hline

Average  &      &     &  2.27E-50 \\ \hline
    St. deviation	   &    &     &  1.96E-51 \\ \hline
		\hline

\end{tabular}
\end{table}

The advantages of using CLASH data is that they provide the values for virial masses M${}_{vir}$ of clusters. Although the degree of virialization of a given cluster varies from one cluster to another and hence is a separate issue for analysis, those masses provide the upper limit of error for the value of $\Lambda$. 

\begin{table}[!h]
\caption{Error limit of $\Lambda$ for CLASH Survey}\label{tab8}
\centering
\begin{tabular}{ |p{3.6cm}|p{1.2cm}|p{2.4 cm}||p{1.8cm}| }
\hline

Cluster& Radius (Mpc) & M${}_{vir} ($ M${}_{\odot}$)&$\Lambda$ ($m^{-2}$) $\leq$\\ \hline
\hline
Abell 383& 1.86  &  (1.04 $\pm {0.07}$)E15 &3.31E-51\\ \hline
Abell 209& 1.95 & (1.17 $\pm {0.07}$)E15 &2.87E-51\\ \hline
Abell 2261& 2.26  &  (1.76 $\pm {0.18}$)E15 &4.74E-51\\ \hline
RXJ2129+0005&1.65  &  (0.73 $\pm {0.07}$)E15 &4.74E-51\\ \hline
Abell 611& 1.79  &  (1.03 $\pm {0.07}$)E15 &3.71E-51\\ \hline
MS2137-2353& 1.89 & (1.26 $\pm {0.06}$)E15 &2.70E-51\\ \hline
RXCJ2248-4431& 1.92  &  (1.40 $\pm{0.12}$)E15 &5.16E-51\\ \hline
MACSJ1115+0129&1.78  &  (1.13 $\pm{0.10}$)E15 &5.39E-51\\ \hline
MACSJ1931-26& 1.61  &  (0.83 $\pm{0.06}$)E15 &4.37E-51\\ \hline
RXJ1532.8+3021& 1.47  &  (0.64 $\pm{0.09}$)E15 &8.62E-51\\ \hline
MACSJ1720+3536& 1.61  & (0.88 $\pm{0.08}$)E15 &5.83E-51\\ \hline
MACSJ0429-02& 1.65  &  (0.96 $\pm{0.14}$)E15 &9.48E-51\\ \hline
MACSJ1206-08& 1.63  &  (1.00 $\pm{0.11}$)E15 &7.73E-51\\ \hline
MACSJ0329-02& 1.54  & (0.86 $\pm{0.11}$) E15 &9.16E-51\\ \hline
RXJ1347-1145& 1.80  &  (1.35 $\pm{0.19}$)E15 &9.91E-51\\ \hline
MACSJ1311-03& 1.28  &  (0.53 $\pm{0.04}$)E15 &5.80E-51\\ \hline
MACSJ1423+24& 1.34  &  (0.65 $\pm{0.11}$)E15 &1.39E-50\\ \hline
MACSJ0744+39& 1.33  &  (0.79 $\pm{0.04}$)E15 &5.17E-51\\ \hline
CLJ1226+3332& 1.57  &(1.72 $\pm{0.11}$)E15 &8.65E-51\\ \hline
\hline

Average  &      &     &  6.38E-51 \\ \hline
    St. deviation	   &    &     &  2.85E-51 \\ \hline
		\hline
 	
\end{tabular}
\end{table}

Namely, by considering the $\Lambda$-modified potential from one side, and the error limit for M${}_{vir}$, on the other side, we have an upper limit for the value of $\Lambda$
\begin{equation}\label{E}
\frac{\mathbb{E}(\sigma^2)}{\sigma^2}= \frac{\Lambda R_{vir}^3 c^2}{6 G M_{vir}},
\end{equation}
where $\mathbb{E}(\sigma^2)$ is the error limit for $\sigma^2$. The results for the CLASH clusters are shown in Table 9. Note that, for all these clusters the error limit covers the value of $\Lambda$ as of cosmological constant.

Meantime, let us note that the virial mass and radius can be defined as \cite{Density}
\begin{equation}\label{VRM}
M=\frac{4}{3} \pi R^3  \Delta_{c} \rho_{crit},
\end{equation}
where $\Delta_{c}$ is the overdensity parameter and $\rho_{crit}$ is the critical density of the universe. The $\Delta_{c}$, in its turn is obtained for each era of the universe as \cite{Galaxy}
\begin{equation}\label{Delta}
\Delta_{c}=18 \pi^2 + 82 x - 39 x^2, \quad x= \Omega(z)-1, \quad \Omega(z)= \frac{H^2_0 }{H^2(z)} \Omega_{0} (1+z)^3,
\end{equation}
where $H$ is the Hubble constant.

Thus it turns out that considering the Newtonian virial theorem $\sigma^2 = \frac{GM_{vir}}{R_{vir}•}$  and Eq.(\ref{VRM}) for virialized systems one can calculate the value of $\Lambda$ based on Eq.(\ref{Es}) 
\begin{equation}\label{Lz}
\Lambda= \frac{3}{4} \frac{\Delta_{c}}{c^2} H^2.
\end{equation}
Eq.(\ref{Delta}) enables one to calculate the value of $\Delta_{c}$ at each era of Universe. For pure de Sitter universe $\Delta_{c} = 18 \pi^2 \approx 178$, while within the ``spherical collapse" model its value yields $\Delta_{c}$ =200 \cite{White}. However, within the $\Lambda$CDM cosmology  $\Delta_{c} \approx$ 100 (\cite{Pace} and references therein). 
Considering all above, for $H=70$ (km/s)/(Mpc) one will obtain $\Lambda \approx 4.30\,\,  10^{-51} m^{-2}$.

\section{Conclusions}

We studied the relevance of the weak field GR with modified potential Eq.(2) to the description of the dynamics of galaxy systems, tracing galaxy pairs, groups and clusters. It is important that the used data are not of the same origin but were obtained at galactic surveys, gravity lensing studies and by Planck satellite. 
For the analysed hierarchy of systems of galaxies the numerical value of $\Lambda$ obtained as a weak field GR without any cosmological considerations is in visible agreement with $\Lambda$ obtained from relativistic cosmology.

There are at least two aspects to be outlined at the interpretation of the obtained $\Lambda$s regarding both the values and their scatter:

a) Eq.(6) used for the estimation of $\Lambda$ assumes that the virial theorem, i.e. the equipartition is determined purely due to the $\Lambda$ term, while obviously the conventional Newtonian term's contribution has to be there as well. Therefore, the cosmological constant Eq.(5) has to be a lower limit for the $\Lambda$ values obtained for galaxy systems. Namely, for fully virialized systems the empirical values of $\Lambda$ have to be close to the cosmological constant, Eq.(5), while for others the empirical values have to be higher. Exactly such a behavior is visible in the exhibited Tables; 

b) the scatter in values of $\Lambda$s in Tables is certainly expected due to the various virialization degree reached in each given system (galaxy pair, group, cluster). Decrease of that scatter will be possible when more refined dynamical structure of any individual system can be available.
     
Thus, while for certain systems e.g. galaxy groups of NGC 2894, NGC 3049, NGC 3810, P 51971 the obtained $\Lambda$s are close to the cosmological constant value, Eq.(5), in their absolute majority they exceed the latter.  

The obtained values of $\Lambda$  are also influenced by the inhomogeneity of the data sources, e.g. by the inevitable difference between the lensing and dynamical masses of galaxy systems or the masses of CMB-SZ clusters (see \cite{DSA}). The structure and the extension of the dark halos of galaxies \cite{G2} are especially relevant for pair galaxies and galaxy groups studies. Note that, a crucial difference of the modified potential Eq.(2) from the Newtonian one is that, the second term in Eq.(2) defines non-force free field inside a shell \cite{G85} and hence e.g. the properties of galactic disks  have to be determined by dark halos \cite{G,G2}. The $\Lambda$-constant can be associated also to the time arrow \cite{AG}. Regarding the $\Lambda$-constant, including in the context of the local Universe within equilibria concepts, see \cite{BT,B,N,N2}. Among other means for probing weak-field modified gravity are the large scale matter distribution (see \cite{E}), the effects in Solar system \cite{KEK,Ciu}.

Thus, the $\Lambda$-constant enables the description both of the accelerated universe and the dynamics of galactic systems. The cosmology is described by GR, while the galaxy system dynamics is described by weak field GR, both containing the same $\Lambda$. One can therefore conclude that, the non-zero cosmological constant might be discovered from galactic systems even before high redshift SN surveys and CMB.  

\section{Acknowledgement}
We are thankful to the referee for valuable comments. AS acknowledges the receipt of the
grant AF-04 from the Abdus Salam International Centre for Theoretical Physics, Trieste, Italy.

\end{document}